\documentclass[10pt, conference, compsocconf]{IEEEtran}
\usepackage{graphicx}

\usepackage{amsmath}
\usepackage{subfigure}
\usepackage{rotating}
\usepackage{multirow}
\usepackage{array}
\usepackage{algorithm}
\usepackage{algorithmic}

\begin{document}

\title{Routing Load of Route Discovery and Route Maintenance in Wireless Reactive Routing Protocols}

\author{D. Mahmood, N. Javaid, U. Qasim$^{\ddag}$, Z. A. Khan$^{\$}$ \\
                $^{\ddag}$University of Alberta, Alberta, Canada.\\
                Department of Electrical Engineering, COMSATS\\ Institute of
                Information Technology, Islamabad, Pakistan. \\
                $^{\$}$Faculty of Engineering, Dalhousie University, Halifax, Canada.}

\maketitle
\begin{abstract}
\boldmath
In this paper, we present analytical study of routing overhead of reactive routing protocols for Wireless Multihop Networks (WMhNs). To accomplish the framework of generalized routing overhead, we choose Ad-Hoc on Demand Distance Vector (AODV), Dynamic Source Routing (DSR) and Dynamic MANET on Demand (DYMO). Considering basic themes of these protocols, we enhance the generalized network models by adding route monitoring overhead. Later, we take different network parameters and produce framework discussing the impact of variations of these parameters in network and routing performance. In the second part of our work, we simulate above mentioned  routing protocols and give a brief discussion and comparison about the environments where these routing protocols perform better.
\end{abstract}


\maketitle

\section{Introduction}

Recent demands of communication make infrastructure communications replaced with infrastructure-less communications. Along with other technologies, WMhNs are promising to provide freedom of communications as they offer supple structures, low costs and ability to cope with ever growing needs of bandwidth. In WMhNs, one node can be out of range with another. Hence to communicate between such nodes, there must be some node/s working as a bridge between them. In other words, intermediate nodes act as router to receive and transmit routing and data packets. This is the reason that each node must work as a routing device. Such networks are gaining popularity day by day. So it is a challenge to maintain and improve quality of WMhNs.

This communication is possible with the help of numerous protocols that are functioning on different layers. Besides other protocols, routing layer protocols play vital role in contributing smoothness and better functionality of a network. A routing protocol creates, maintains and synchronizes a routing table and relevant routing information for a node. More efficient routing layer protocol results in more efficient network performance. Numerous protocols have been developed and can be categorized into two major classes i.e., reactive routing protocols and proactive routing protocols. Reactive routing is based on the immediate response phenomena [1]. In our work we emphasis on reactive routing only. It is a common observation that lots of simulated work has been contributed with respect to analytical modeling of above mentioned routing protocols. This was the basic motivation to work on mathematical framework that gives and discusses precise information about the behaviors of reactive routing protocols in different environments and parameters. For this purpose, we choose three main routing protocols from reactive routing i.e.,  DYnamic MANET On-demand $(DYMO)$ [2], Ad-hoc On-demand Distance Vector $(AODV)$ [3] and Dynamic Source Routing $(DSR)$ [4]) for our studies.

\section{Related Work and Motivation}
Authors in [5] provide analytical framework for calculating routing overhead of reactive protocols. They quantify route discovery process, i.e., overhead due to route REQuest packets and route REPly packets of any network underlying a reactive routing protocol. However, link monitoring overhead is not considered in their work. [6] gives a combined framework of reactive and proactive routing protocols. The proposed models express scalability issues of a network considering both classes of routing protocols i.e., reactive and proactive. In [7], authors propose analytical model which presents the effect of traffic on routing overhead whereas, [8] presents a survey of routing overhead on both reactive and proactive protocols and discuss cost of energy as routing metric. 
 I.D Aron \emph{et.al} [9] present link repairing modeling, both in local repairing and source to destination repairing of two routing protocols, which were DSR and WRP. They compare these two routing protocols, though aggregate routing overhead is not considered in [9]. In [10], authors present brief understanding of scalability issues of network, however, impact of topology change is not sufficiently addressed.

We enhance the framework produced by [5] by adding link monitoring overhead and trigger message overhead. Hence, we present a general routing overhead framework. After calculating the aggregate routing overhead, we calculate rate of change in different network and protocol parameters. Besides modeling, we simulate above mentioned three reactive routing protocols and discuss their behaviors according to different environments and scenrios.

\section{Modeling Routing Operations}
In our work, we assume that nodes of network are placed in grid environment while nodes have different life times. Certain sections of the grid are vulnerable to failure due to some reasons. That can be the power failure or radio jamming. If one section of grid fails for certain time, it surely gives variations in number of nodes, number of hops and obviously in link monitoring overhead.

\subsection{Reactive Route Discovery Overhead}
Route discovery overhead can further split into two parts i.e., Overhead due to Route REQuest (RREQ) dissemination and  Overhead due to generation of Route REPly (RREP). Considering RREQ dissemination Overhead, it mainly depends upon the number of hops a packet travels and number of neighbors of nodes at each hop.

\begin{figure}[t]
\begin{center}
  \includegraphics[height=6cm, width=6cm]{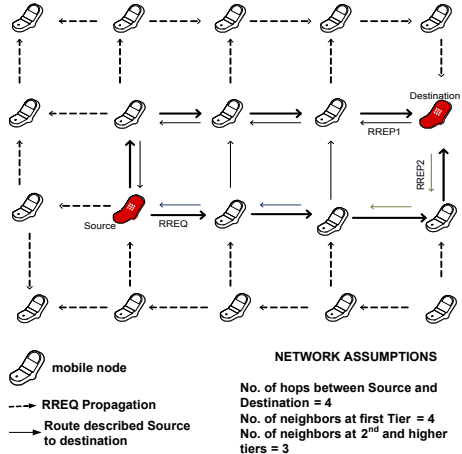}
  \vspace{-0.4cm}
  \caption{Propagation of RREQ and RREP in Network}
\end{center}
\end{figure}

Route Request is propagated in entire network till destination is find. If we consider that source and destination nodes of a network are placed at opposite corners of network, then they bear highest number of hops [11]. As shown in Fig. 1, when a RREQ is generated on the originator node, there are four neighbors, though, from second tier up to $n$ tiers,  number of effective neighbors are three. Moreover, at each intermediate node, there is some coverage index where the packet is processed [12]. As, RREQ packet is blind flooded that allows a node to process RREQ control packet once and discard itself if receives second time. [5] presents the mathematical model for route discovery overhead i.e., RREQ overhead and RREP overhead of reactive routing. According to this mathematical modeling, overhead of RREQ is:

\tiny
\begin{eqnarray}
R_{RREQ}&=&\sum_{n-1}^{H} (4)3^{H-1}\sum_{i=2}^{4}[(n-1-i)-\sum_{j=1}^{H-1}N_{j}]pC_{i}
\end{eqnarray}
\normalsize

$C\emph{i}$ is the additional coverage index of a node having $i$ nodes as its neighbors. $C\emph{i}$ is described and presented in [12], $H$ is the expected number of hops of network and $N_j$ is the expected number of neighbors at $\j^{th}$ hop.

Once RREQ reaches destination, RREP is generated. It follows the reverse path. For simplifying the concept, in Fig. 1, two route RREP's are generated i.e., two paths to destination are found. Considering the same point of view, [5] has given the expected overhead generated by RREP packets as:

\small
\begin{eqnarray}
R_{RREP}&=&H+ \frac{H}{2}(n-h-2)p
\end{eqnarray}
\normalsize
Overall routing overhead due to route discovery can be represented as:

\small
\begin{eqnarray}
R_{DISCOVERY}&=&RREQ+RREP
\end{eqnarray}
\normalsize
Placing values in $Eq. 3$ we get

\tiny
\begin{align}
R_{DISCOVERY}&=&\sum_{n-1}^{H} (4)3^{H-1}\sum_{i=2}^{4}[(n-1-i)-\sum_{j=1}^{H-1}N_{j}]pC_{i}+ \frac{H}{2}(n-h-2)p
\end{align}
\normalsize
\subsection{Reactive Route Maintenance Overhead}
\begin{figure}
  \begin{center}
  \includegraphics[height=6cm, width=6cm]{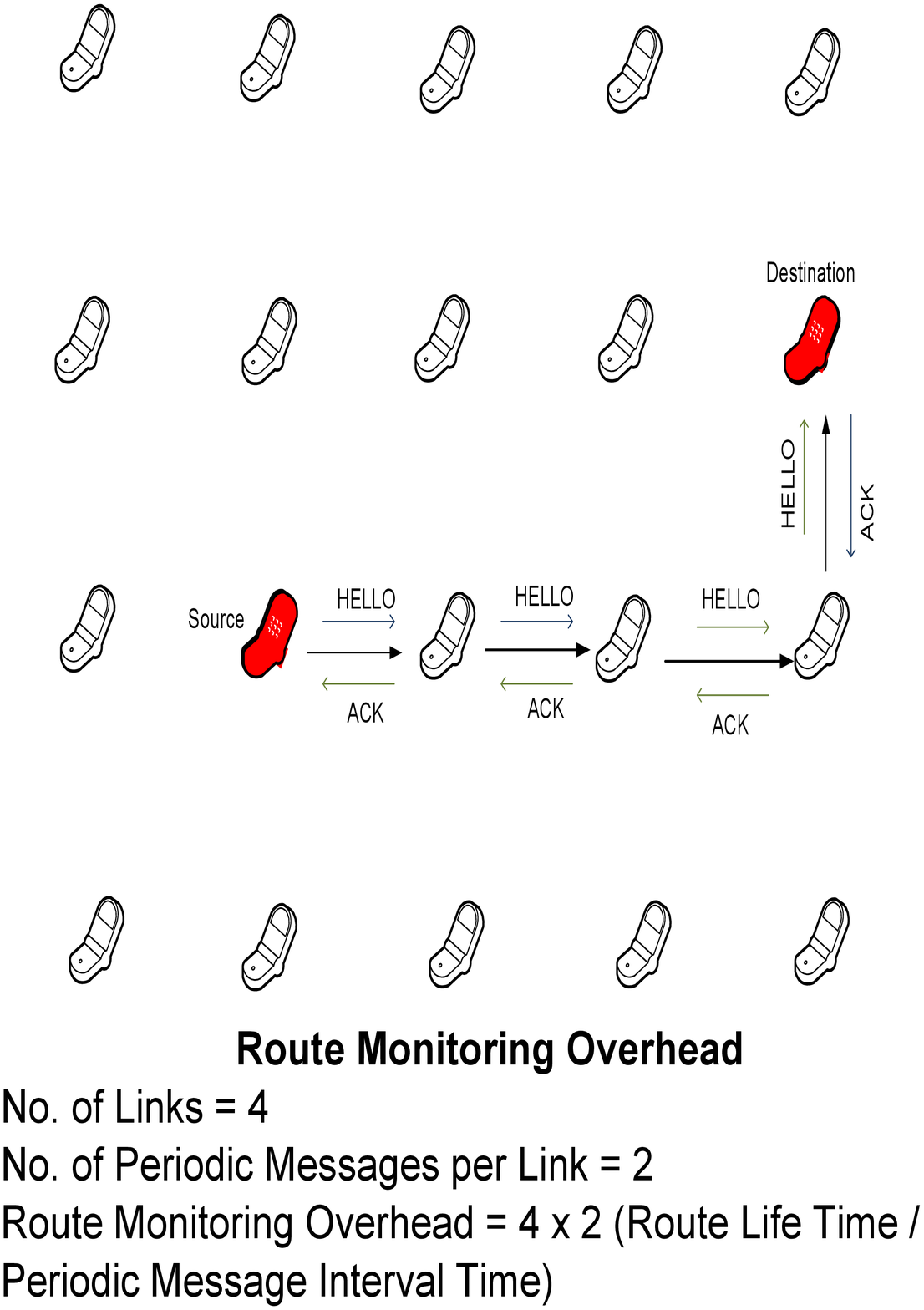}
    \vspace{-0.4cm}
  \caption{Link Monitoring Overhead of a Route}
  \end{center}
\end{figure}

Link monitoring overhead in reactive protocols is minor with respect to route discovery overhead, however, it also contributes in aggregate routing overhead. $Hello$ messages are periodically propagated when a route is established till expiration of route in AODV. In DSR $ACK$ message works almost in the same fashion [3-4]. As we know that life time of any link in a route is a random variable. It can be defined as the time when two nodes create a pair with each other till the time, this pair is prone to breaks due to any reason mostly due to radio problems [13].\\
As depicted in Fig. 2 , when a route is established having $l$ links, a link monitoring message is propagated periodically till the route expires. Hence, total number of hello messages broadcasted for monitoring of one route whose route expiry time is $T$ and periodic interval time is $t$ can be represented mathematically as:

\small
\begin{eqnarray}
R_{HELLO(e)}&=&2(\frac{T}{t})l
\end{eqnarray}
\normalsize
And number of periodic link monitoring messages for $n$ routes can simply be represented as:

\small
\begin{eqnarray}
R_{HELLO}&=&\sum_{i=1}^{n}2(\frac{T}{t})l
\end{eqnarray}
\normalsize
\subsection{Aggregate Reactive Overhead}
Overall routing overhead due to route discovery and route monitoring is be stated as:

\small
\begin{eqnarray}
RO&=&R_{DISCOVERY}+R_{HELLO}
\end{eqnarray}
\normalsize
Putting values from Eq.4 and Eq.6, we get:

\tiny
\begin{eqnarray}
RO&=&\sum_{n-1}^{H} (4)3^{H-1}\sum_{i=2}^{4}[(n-1-i)-\sum_{j=1}^{H-1}N_{j}]p(_{i})+H+ \frac{H}{2}(n-h-2)p+\nonumber\\
& &\sum_{i=1}^{n}2(\frac{T}{t})l
\end{eqnarray}
\normalsize
One can not make bricks without clay. In proposed network model, nodes are placed in a grid that works as clay for calculating reactive routing overhead. However, if we consider that these nodes have different life times within the same network, or have power failure at certain sections of grid then number of hops as well as number of nodes, at that instance, can be varied. To calculate overhead during such scenarios, we can take partial derivatives with respect to number of hops and number of nodes of network. Besides number of hops and number of nodes of network, we take parameters of routing protocols as route life time and periodic messages interval time which can also be varied. Hence, we have two more metrics while modeling routing overhead.

\begin{figure}
\begin{center}
\includegraphics[height=6cm, width=6cm]{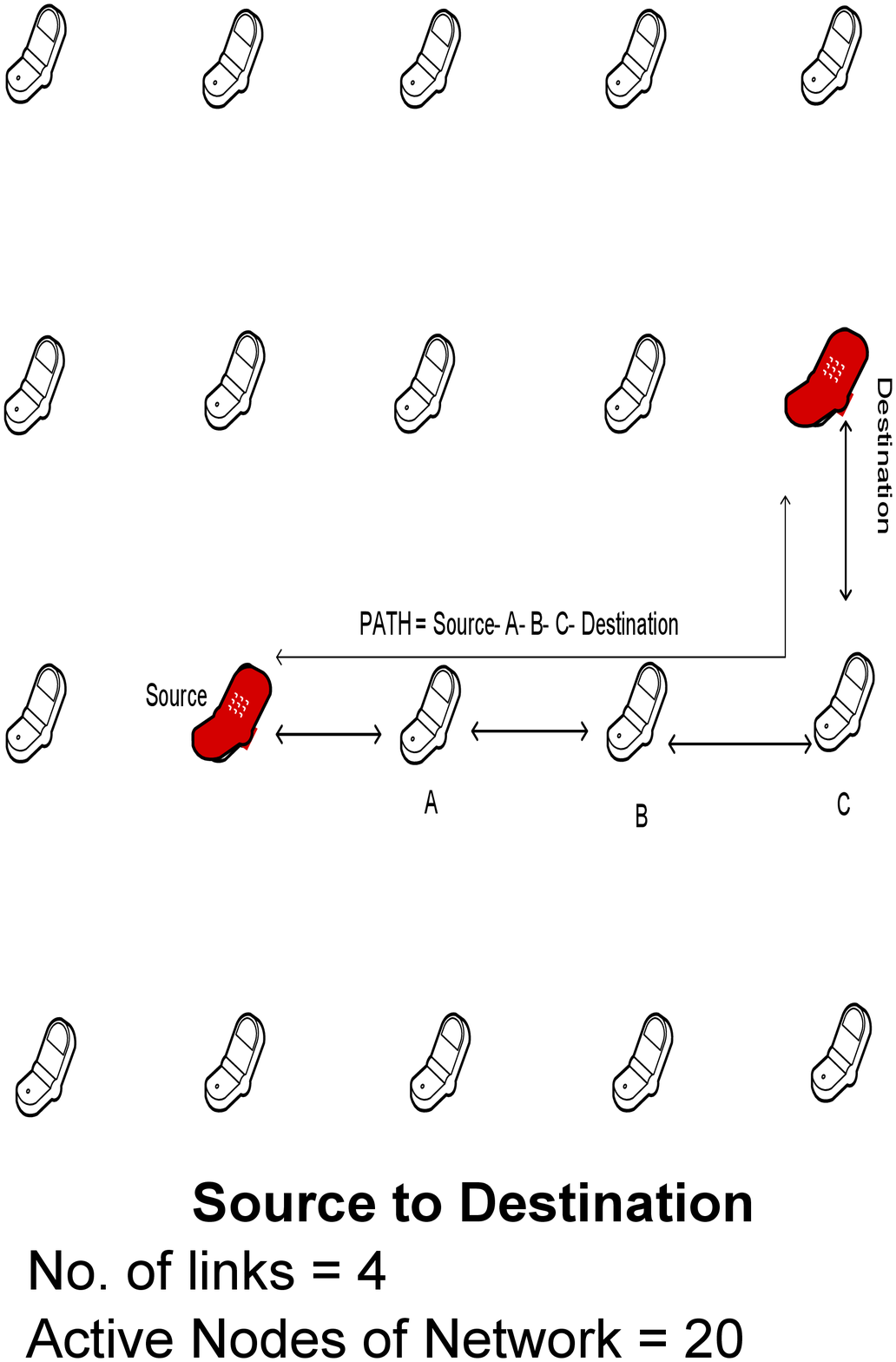}
\vspace{-0.4cm}
\caption{Source to Destination Route(Grid Environment)}
\end{center}
\end{figure}

\begin{figure}
\begin{center}
\includegraphics[height=6cm, width=6cm]{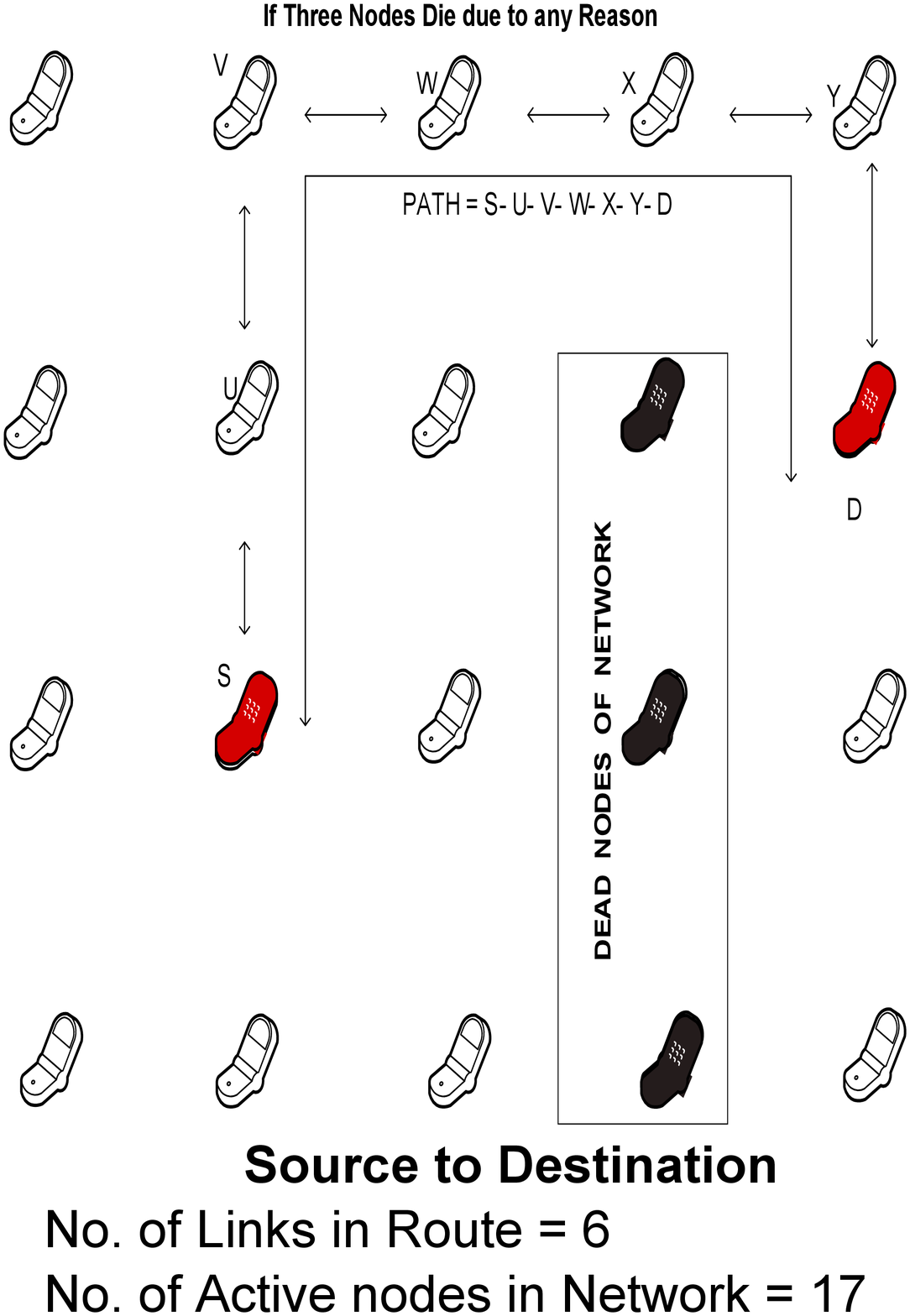}
\vspace{-0.4cm}
\caption{Source to Destination Route(Portion of Grid Black Out)}
\end{center}
\end{figure}

Considering Fig. 3, $Eq. 8$ gives the routing overhead of reactive routing protocol. However, if there is some change in the number of nodes of network due to any reason like power failure or radio jamming, the source requires a new route to its destination, as illustrated in Fig. 4. That new route bears different number of intermediate nodes as well as hops or links. To calculate such variations in the network, we have to analyze rate of change in routing overhead with respect to different parameters.

From $Eq. 8$, function $y$ (i.e. $RO$) we get different parameters as $n$, $H$, $T$ and $t$. As discussed earlier, $n$ stands for total number of nodes in a network, $H$ represents number of hops of a network, $T$ is the route life time while $t$ is periodic interval time for link monitoring. If we take partial derivative with respect to number of nodes of network, periodic interval time, route life time and number of hops of network, we get to know the overall rate of change in the network with respect to routing overhead.

\tiny
\begin{eqnarray}
y (n,H,T,t)&=&\sum_{n-1}^{H} (4)3^{H-1}\sum_{i=2}^{4}[(n-1-i)-\sum_{j=1}^{H-1}N_{j}]p.C_{i})+\nonumber\\
& &H+ \frac{H}{2}(n-h-2)p
\end{eqnarray}
\normalsize
Considering the parameters of function $y$, it is interesting to know that number of hops or links are dependent on the number of nodes of a network. Whereas, number of link monitoring messages are dependent upon number of links of route, link life time and periodic message life time. Their relationships are discussed further in the coming equations. Considering the variation in number of nodes, we take partial derivative of $y$ with respect to $n$ and we get:

\tiny
\begin{eqnarray}
{\partial{y}}/{\partial{n}}&=&\sum_{n-1}^{H} (4)3^{H-1}[\sum_{i=2}^{4}[(-i)-\sum_{j=1}^{H-1}N_{j}]p.C_{i}]+\nonumber\\
& &H+ \frac{H}{2}(-h-2)p
\end{eqnarray}
\normalsize
Varying number of nodes of a network certainly effects the number of hops of some routes. Such variation can result in change in routing overhead. Number of hops are in relationship with the number of links per route and number of links is a vital parameter for link monitoring overhead. When number of nodes  vary, there is a possibility of change in number of hops. To calculate this change, we take partial derivative of function $y$ with respect to $H$.

\tiny
\begin{eqnarray}
{\partial{y}}/{\partial{H}}&=&\sum_{n-1}^{H} (4)3^{H-1}+ H-1(3^{H-1})[\sum_{i=2}^{4}[(n-1-i)-\sum_{j=1}^{H-1}N_{j}]p.C_{i}]+\nonumber\\
& &1+ \frac{1}{2}(n-3)p
\end{eqnarray}
\normalsize

Number of nodes and number of hops play a vital role in routing overhead. We can use chain rule to calculate overall routing overhead, assuming route life time and periodic link monitoring message interval constant. For this purpose, we have our $Eq. 4$ discussing routing overhead due to route discovery process. Considering $Eq. 4$ as a function $x$ we get our partial derivatives of $n$ and $H$ as $Eq. 10$ and $Eq. 11$. Routing overhead due to varying number of nodes and number of hops can be calculated as:
\small
\begin{align}
dx =(\frac{\partial{x}}{\partial{n}})dn+(\frac{\partial{x}}{\partial{H}})dH
\end{align}
\normalsize
placing values in Eq.12, we get::

\tiny
\begin{eqnarray}
dx&=&(\sum_{n-1}^{H} (4)3^{H-1}[\sum_{i=2}^{4}[(n-1-i)-\sum_{j=1}^{H-1}N_{j}]p(C_{i})]+H+ \frac{H}{2}(n-h-2)p)dn+\nonumber\\
& &(\sum_{n-1}^{H} (4)3^{H-1}[\sum_{i=2}^{4}[(n-1-i)-\sum_{j=1}^{H-1}N_{j}]p(C_{i})]+1+ \frac{1}{2}(n-3)p)dH+\nonumber\\
\end{eqnarray}
\normalsize
Considering rate of change in route life time:

\small
\begin{eqnarray}
{\partial{y}}/{\partial{T}}&=&\sum_{i=1}^{n} (\frac{2}{t})l_{i}
\end{eqnarray}
\normalsize
And to analyze variation in periodic interval time for link monitoring:

\small
\begin{eqnarray}
{\partial{y}}/{\partial{t}}&=&\sum_{i=1}^{n} -2(\frac{T}{t^{2}})l_{i}
\end{eqnarray}
\normalsize
Considering $Eq. 14$ and $Eq. 15$, we can conclude that if there are different route life times active in a network along with different periodic message intervals, than overall routing overhead of link monitoring of a Reactive Routing Protocol is the total derivative with respect to route life time and periodic message interval. To calculate so, we consider $R_{HELLO}$ (expressed in $Eq. 6$)as a function $z$ whose partial derivatives are expressed in $Eq. 14$ and $Eq. 15$. Taking total derivative we get:

\small
\begin{eqnarray}
dz =(\frac{\partial{z}}{\partial{T}})dT+(\frac{\partial{z}}{\partial{t}})dt
 \end{eqnarray}
\normalsize
Placing the values we get the routing overhead due to varying routing protocol parameters of route life time and periodic message update time:
\small
\begin{eqnarray}
dz&=&
(\sum_{i=1}^{n} (\frac{2}{t})l_{i})dT
 + (\sum_{i=1}^{n} -2(\frac{T}{t^{2}})l_{i})dt
\end{eqnarray}
\normalsize

\normalsize
Applying chain rule on the function $y$ to get the total derivative which is actually the sum of all the partial derivatives of a function, we get the optimum model for route discovery and route monitoring overhead in reactive routing protocols.

\small
\begin{align}
dy =(\frac{\partial{y}}{\partial{n}})dn+(\frac{\partial{y}}{\partial{H}})dH+(\frac{\partial{y}}{\partial{T}})dT+(\frac{\partial{y}}{\partial{t}})dt
\end{align}
\normalsize
Placing the values, we get:

\tiny
\begin{eqnarray}
dy&=&(\sum_{n-1}^{H} (4)3^{H-1}[\sum_{i=2}^{4}[(n-1-i)-\sum_{j=1}^{H-1}N_{j}]p(C_{i})]+H+ \frac{H}{2}(n-h-2)p)dn+\nonumber\\
& &(\sum_{n-1}^{H} (4)3^{H-1}[\sum_{i=2}^{4}[(n-1-i)-\sum_{j=1}^{H-1}N_{j}]p(C_{i})]+1+ \frac{1}{2}(n-3)p)dH+\nonumber\\
& &(\sum_{i=1}^{n} (\frac{2}{t})l_{i})dT
 + (\sum_{i=1}^{n} -2(\frac{T}{t^{2}})l_{i})dt
\end{eqnarray}
\normalsize

\section{Simulation Results and Discussions}

We use NS-2 as our simulation tool. AODV [15] coding was developed by CMU/MONARCH group while it was optimized by Samir Das and Mahesh Marina (University of Cincinnati). Coding of DYMOUM by MASIMUM [17] is used for DYMO. We use NS-2.34 for simulating AODV and DSR while, DYMOUM is simulated in NS-2.29. We focus on the mobility and scalability factors of Ad Hoc networks in our work.

We considered a network of 50 nodes where nodes are randomly located and are mobile. These nodes have a bandwidth of $2\,\,Mbps$ each. Mobility is set as $2\,\,m/s$ which is average walking speed. Packet size is defined as $512\,\,bytes$, while simulation setup runs on Continues Bit Rates (CBR). The size of network is defined as $1000\,\,m^2$.
Given these parameters, we have confined our experiments to following three metrics.\\
1.	Throughput\\
2.	End to End Delay\\
3.	Normalized Routing Load.

\subsection{Throughput of Reactive Protocols}
\begin{figure}
\begin{center}
\includegraphics[scale=0.4]{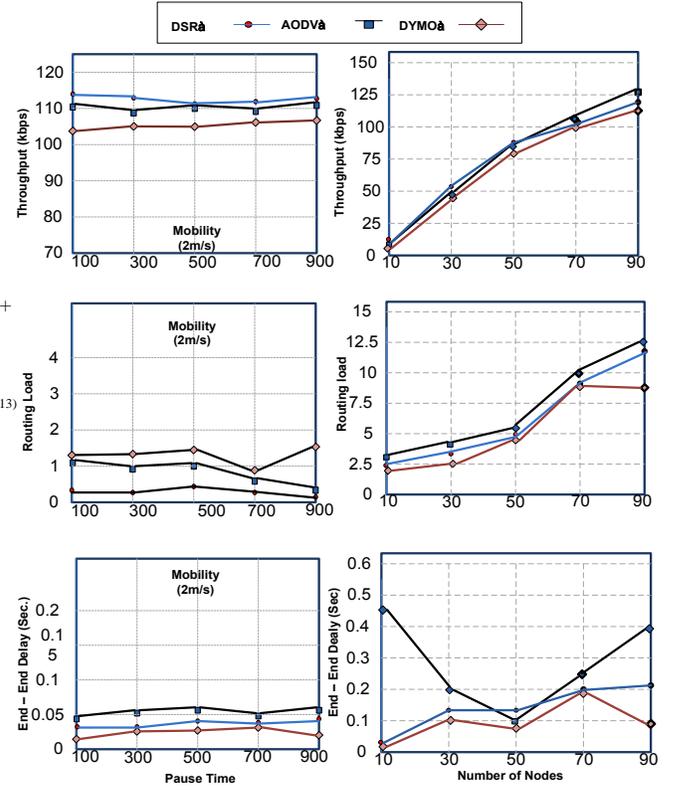}
\caption{Simulation Results of Reactive Protocols: AODV, DSR, DYMO}
\end{center}
\end{figure}

In general sense, throughput refers to the amount of data that has successfully reached its destination. Mathematically it can be stated as:

\begin{equation}
Througput=\frac{messages Recieved Successfully}{Time}
\end{equation}

\emph{Mobility Factor:} Considering graphs for throughput, DSR attains the maximum throughput with respect to AODV and DYMO. If we consider AODV, than it, surely have a $TIMEOUT$ factor involved. AODV waits for a specified time, then route is termed invalid and finally erased from routing table. ``HELLO'' messages (used for link monitoring) in AODV also works very well for mobile environment. Overall considering mobility factor, DSR gives stable throughput, as no unnecessary packets are generated by this routing protocol. In link breakages, DSR have multiple routes while, in AODV, routing table keeps the best chosen path only. Hence, within the environment where links are immune to breaks, DSR supersedes AODV and DYMO.  DYMO proves to be the worst amongst the other two protocols.\\

\emph{Scalability Factor:} According to experiments performed, AODV converges at almost all data rates with salabilities. While DSR proves itself to be scalable but only during high data traffic, it can not converge the network. DYMO performs worst among these studied routing protocols. As the number of nodes increases or data traffic increases, its performance degrades dramatically. According to [14], a network of multiple thousands of nodes with different traffic loads can be handled by AODV. The reason that AODV supersedes DSR and DYMO is lower packet loss ratio and propagation of information regarding distant vector which practically consume minimum bandwidth. This feature gives AODV a room for scalability. In AODV, routing packet contains only one hop information while in DSR, packet size is larger as it keeps the information of whole route. This is another reason that AODV outperforms DSR.

\begin{table}
\centering
\caption{Comparison: Reactive Routing Protocols}
  \begin{tabular}{p{1.5cm}| p{1.5cm} | p{1.5cm} | p{1.5cm}  }
  \hline{}
    \textbf{Feature}          & \textbf{AODV}    &   \textbf{DSR}   & \textbf{DYMO}	       \\
\hline{}
Protocol type  & Distance Vector   &	Source routing   & Source routing	\\
\hline{}
Route maintained in	& Routing table &	Route Cache	      & Routing table     \\
\hline{}
Multiple route discovery &	No       &	Yes               &	No	                \\
\hline{}
Update destination & Source	        & Source	           & Source	            \\
\hline{}
Broadcast	      & Full	         & Full	                 & Full	            \\
\hline{}
Reuse of routing information	& No	& Yes	& No	 \\
\hline{}
Route selection & Only searched route	& Hop count	      & Only searched route	   \\
\hline{}
Route reconfiguration & Erase route notify source & Erase route notify source	& Erase route notify source 	\\
\hline{}
Route discovery packets	& using RREQ and RREP packets	& using RREQ and RREP packets	& using RREQ and RREP packets	 \\
\hline{}
Limiting overhead, collision avoidance, network congestion	& Expanding Ring Search Algorithm 	& Expanding Ring Search Algorithm	& Expanding Ring Search Algorithm   \\
\hline{}
Limiting overhead, collision avoidance, network congestion	& Binary Exponential Back off Time	& Binary Exponential Back off Time &	Binary Exponential Back off Time	 \\
\hline{}
Update information	& By RERR message	& By RERR message	& By RERR message	 \\
\hline{}


  \end{tabular}

\end{table}
\subsubsection{End to End Delay of Reactive Routing}
 Time which a packet takes in reaching destined node from the originator node can be termed as end to end delay. Mathematically we can express it as:\\

$ED=\frac{(Number of Transmitted Packets)(RTT)}{Number of Recieved Packets}$\\

\emph{Mobility Factor:} As shown in the graphs, AODV gives lowest performance as, link breakages may lead to longer routes. DYMO, though works worst in throughput case but here it works best amongst DSR and AODV. It is so because, DYMO does not check the routes in memory as DSR looks into route cache and AODV in to its routing table, instead it starts Expanding Ring Search $(ERS)$ algorithm whenever a route is required.\\

\emph{Scalability Factor:} The concept of $gratitous$ RREP is used both in DSR and AODV. This is the reason that DYMO results in lowest End to End delay, irrelevant of number of nodes in the network. $Gratitous$ RREP though results in lower delay at normal traffic rates though, DSR checks the route cache before starting Expanding Ring Search (ERS) algorithm in the same way as AODV search route in its routing table before starting a route request using ERS Algorithm. DYMO does not use such stored information rather it simply initiates ERS.  AODV also have a link repair feature that makes it bear the highest end to end delay with respect to any scalability among DYMO and DSR.\\

\subsubsection{Routing Load of Reactive Routing}
When a single data packet is to be sent from one node to another within a network, a number of routing packets are involved in sending this data packet. The numbers of these routing packets which are sent just to transfer one data packet are termed as Routing Load or Normalized Routing Load. Mathematically, we can state:\\
$ Routing Load = (Routing + Data Load) - (Number of Data Packets sent)$\\

\emph{Mobility Factor:} AODV and DSR use the concept of $grat.$  RREP, i.e. when a $RREQ$ reaches any node that has a valid route stored in its route cache or routing table, it generates a $RREP$ by itself to the original source node. This $RREP$ contains the full information up to the destined node and overhead of finding route beyond that node limits. DYMO does not use this $grat.$ RREP. That's why it suffers from greater routing overhead with respect to the other two protocols. AODV also works well in the context of normalized routing overhead however, there is a concept of local link repair and above all, use of $HELLO$ message for link monitoring, makes it performance lower then DSR. A node with underlying DSR protocol use promiscuous mode and this is the reason that it bears lowest overhead.\\
A common observation with respect to increase in mobility of nodes in the network is that all the three routing protocols bear gradually higher overhead. The reason is propagation of route error packets. As the mobility increases, chances of link breaks also increase in the same proportion which results in increase of routing overhead.

\emph{Scalability Factor:} Routing overhead of DYMO is lower than that of AODV and DSR. AODV bears high routing overhead in dense networks. Periodic link sensing packets involved in local link repair mechanism and $grat. RREP$ results in high routing overhead. Whereas promiscuous mode utilized by DSR reduces the routing overhead in not so dense environment.

\subsection{Discussion}
The protocol that uses minimum resources of bandwidth by its control packets can provide better data flow. Hence, the environments where traffic load is very high, protocols having low routing overhead survive. If we consider scalability, than AODV stands at top of rest of studied routing protocols. It uses distance vector distribution that minimize network resource consumption. The network underlying AODV protocol bears low routing overhead as control packets of AODV contains a very small part of information in them where as if we compare it with DSR, control packet of DSR carries whole routing information in it. Hence we can say that DSR has higher routing overhead in terms of bytes or size. If we consider number of control packets than DSR broadcast less number of packets than that of AODV. AODV use periodic hello packet for link sensing and also bear local repair routing overhead. Hence if we compare both of these routing protocols (AODV and DSR) considering mobility and speed factors, we can conclude that both of these protocols give more or less same performance.

Concluding all the routing protocols, our study suggest that, AODV can be selected for denser environments where lower routing overhead is required, DSR should be used within a network having limited number of hops but it is better for highly mobile environment. DYMO routing protocol can be used in networks where delay is in tolerable. As like other reactive protocols, DYMO does not look for any stored route as DSR looks into its cache and AODV in its routing table. It initializes binary exponential back off and ERS algorithm immediately.

\section{Conclusion}
In our work, we have discussed and presented generalized routing overhead framework of reactive routing protocols in WMhNs. We analyze route discovery overhead plus route monitoring overhead. Furthermore, we presented framework with variation in number of nodes of a network, number of hops of a network, route life time of a route, periodic update interval time and frequency of trigger updates to better understand the behavior and functionality of routing protocol. We have discussed overhead due to RREQ packets, RREP packets and link monitoring. In future, we will analyze the routing overhead due to link repair processes as well.
Later experiments are conducted on three routing protocols keeping parameters of throughput, End to End delay and routing overhead in emphasis with respect to mobility and scalability factors. These experiments shows that AODV performs better in highly scalable environment where as DSR works almost in same fashion as of AODV in mobility scenario. Though DSR performs better if there are less number of hops with in the network.

\end{document}